\documentclass{article}
\usepackage{spconf,amsmath,graphicx}
\usepackage{multirow}
\usepackage[psamsfonts]{amssymb}
\usepackage{amsxtra}
\usepackage{threeparttable}
\usepackage{makecell}
\usepackage{hyperref}
\usepackage{url}
\PassOptionsToPackage{hyphens}{url}

\title{Two-stage training method for Japanese electrolaryngeal speech enhancement based on sequence-to-sequence voice conversion}
\name{Ding Ma, Lester Phillip Violeta, Kazuhiro Kobayashi, Tomoki Toda}
\address{Nagoya University, Japan}
\copyrightnotice{978-1-6654-7189-3/22/\$31.00~\copyright2023 IEEE}
\begin{document}
%
\maketitle
\begin{abstract}
Sequence-to-sequence (seq2seq) voice conversion (VC) models have greater potential in converting electrolaryngeal (EL) speech to normal speech (EL2SP) compared to conventional VC models. However, EL2SP based on seq2seq VC requires a sufficiently large amount of parallel data for the model training and it suffers from significant performance degradation when the amount of training data is insufficient. To address this issue, we suggest a novel, two-stage strategy to optimize the performance on EL2SP based on seq2seq VC when a small amount of the parallel dataset is available. In contrast to utilizing high-quality data augmentations in previous studies, we first combine a large amount of imperfect synthetic parallel data of EL and normal speech, with the original dataset into VC training. Then, a second stage training is conducted with the original parallel dataset only. The results show that the proposed method progressively improves the performance of EL2SP based on seq2seq VC.

\end{abstract}

\setlength{\abovecaptionskip}{-0.5cm}
\begin{figure*}[!t]
\centering
\includegraphics[width=140mm]{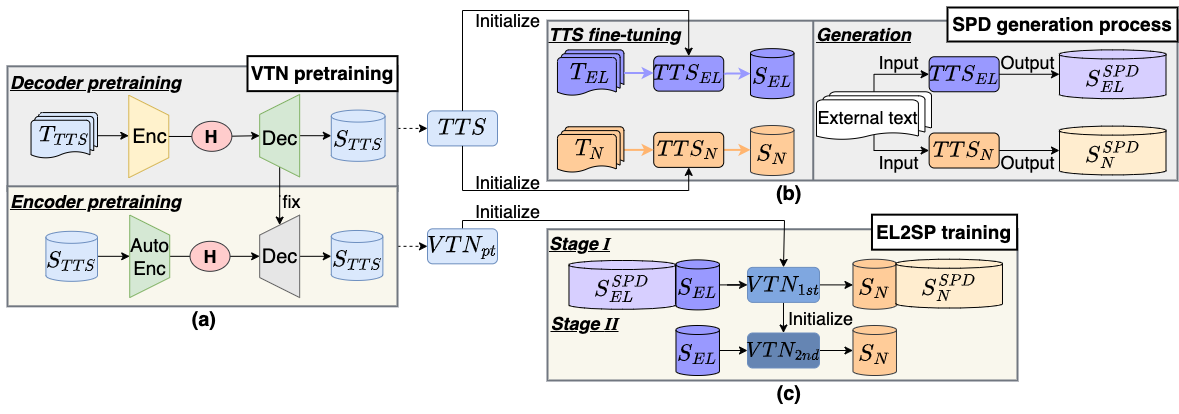}
\vspace*{-5pt}
\caption{The overall system for improving seq2seq-based EL2SP enhancement.} 
\label{Fig. 1}
\vspace*{-10pt}
\end{figure*}

\section{Introduction}
\label{sec:intro}

People suffer from laryngeal cancer or accidents have to undergo laryngectomy to partially or totally remove the laryngeal organs, including the vocal folds. As a result, they rely on a portable medical device named electrolarynx to produce electrolaryngeal (EL) speech as an alternative speaking method \cite{1}. However, EL speech faces two major problems: 1) High energetic excitation signals are always accompanied by radiated noise, which significantly degrades the quality of EL speech and causes distress to the surrounding people. 2) Unnatural speech as the excitation signals provided to the vocal tract cannot simulate the variable F0 contours of the natural human voice. These problems can lead to significant impairment on qualities of patients' lives, and hence urgently need to be addressed.

Voice conversion (VC) is originally designed to convert the speaker identity of speech from source speaker into target speaker while preserving the linguistic information \cite{2}, and it can also be applied as an enhancement to the EL speech to normal speech conversion (EL2SP). Many researches of EL2SP are based on conventional statistical models \cite{3, 4}, which follows a frame-to-frame mapping paradigm, i.e., the converted speech must have the same temporal structure as the source EL speech. This is detrimental for the EL2SP performance in particular prosodic feature conversion.

Compared with conventional VC models, sequence-to-sequence (seq2seq) models \cite{5} are more powerful and flexible to address the difficulties encountered by EL2SP. Seq2seq VC models employ an attention-based encoder-decoder framework to automatically determine the output duration patterns of synthesized speech. Moreover, seq2seq models can capture long-term dependencies such as prosody, suprasegmental characteristics of F0 and speaker identity \cite{6}. Thus, seq2seq VC is promising for optimizing the information loss and misalignments in EL speech. A few studies \cite{7, 8} demonstrated the potential of seq2seq models in speaking-aid, the latter also showed their advantages over non-seq2seq ones. However, most seq2seq models require a huge parallel training corpus. The issues such as mispronunciations and skipped phonemes often occur especially when training data is insufficient \cite{9}.

To address this issue, the seminal work by \cite{10} used highly intelligible synthetic voice from the text-to-speech (TTS) model to implement a many-to-one VC and applied it to hearing-impaired speech recovery. Subsequently, a seq2seq model named Voice Transformer Network (VTN) \cite{6} was extended with synthetic parallel data (SPD) to tackle non-parallel datasets \cite{11}. Following that, \cite{12} investigated the effectiveness of SPD in VC between normal speakers. In light of this, we propose a novel, two-stage method to optimize the seq2seq model for more challenging EL2SP with SPD. In contrast to the aforementioned strategy in \cite{10}, we manage to use SPD with much lower quality since it is impractical to build high-performance TTS models based on the low-resource dataset of EL2SP. We investigate the feasibility of the EL2SP method and the effectiveness of SPD. The main contributions of the study are as follows.

\begin{itemize}
\setlength{\itemsep}{0pt}
\setlength{\parsep}{0pt}
\setlength{\parskip}{0pt}
\item Comparing to the majority systems having demanding requirements for the data, we propose an efficient and robust EL2SP system based on low-quality SPD.
\item We show that the proposed system is able to outperform a typical pretrained seq2seq VC model in terms of naturalness and intelligibility with a small and even semi-parallel original dataset.
\item The feasibility of SPD, which is an easy-to-access but unclean data, on EL2SP is verified.
\end{itemize}

\section{Related works}
\subsection{Sequence-to-sequence voice conversion }
\label{2}

Seq2seq VC model can encode an input source speech $\textbf x_{1:n} = \{ \textbf x_{1}, \textbf x_{2} ,...,  \textbf x_{n} \}$ into a sequence of hidden representations $\textbf h_{1:n}=\{ \textbf h_{1}, \textbf h_{2},..., \textbf h_{n} \}$. During the autoregressive decoding process at each time step, the attention mechanism considers the hidden representations of the encoder outputs at all previous time steps and then generates the context vector. After that, the decoder predicts the output acoustic features of the context vector and finally generates the converted speech $\textbf y_{1:m} = \{ \textbf y_{1}, \textbf y_{2} ,..., \textbf y_{m} \}$ by employing post-network optimization \cite{13}. Based on the above mechanism, the seq2seq VC model can relax the alignment restriction to realize the mapping between different lengths, i.e., $n$$\neq$$m$.

A plethora of works on constructing seq2seq VC are comprised of either recurrent neural networks (RNNs) \cite{14} or convolutional neural networks (CNNs) \cite{15}. In general, RNNs have difficulties in discovering effective long-term dependencies due to the properties of recursive layers, while CNNs rely on the number of the different kernels in convolutional layers, particularly in handling long sequences. Conversely, having great potential in voice processing, the Transformer \cite{16}, possesses multi-head self-attention mechanism and positional embedding, making it capable of perceiving global inputs efficiently. In our work, we find out that EL2SP can inherit the hidden representations of normal-speaker VC. Inspired by \cite{8}, the system we built is able to share the hidden representations from Transformer-based TTS model to the VC model through pretraining.

\subsection{Data augmentation in voice conversion}
Data augmentation is an efficient approach to increase the variety and quantity of the dataset by generating synthetic training data, thus resolving the issues of non-parallel, semi-parallel, or limited data in VC \cite{11}. \cite{17} presented a data augmentation method named ParaGen for frame-to-frame non-parallel VC, but required a sophisticated architecture with accurate hyperparameters and costly training for speaker disentangling to produce the augmented speech. And \cite{10} proposed a VC system named Parratron to normalize arbitrary speaker's speech into a single canonical target speech by leveraging synthetic data. However, building the TTS model relies on huge amount of data. By and large, most studies focus on generating high-quality augmented data. In our study, large amount of original dataset is assumed unavailable. The trained TTS models are directly implemented to produce relatively low-quality SPD for VC.

\section{Proposed method}
\label{3}

The central idea of the proposed method is to improve the performance of the seq2seq EL2SP system, while only requiring a small number of initial dataset even with the presence of non-parallel parts, and its provided text. As explained in Fig. \ref{Fig. 1}, the whole system consists of three steps: (a) we first conduct the pretraining of Voice Transformer Network (VTN) using a large dataset of normal corpus (see Section 3.1); (b) then we fine-tune the Transformer-based TTS model by using a small original dataset to generate SPD for EL and normal speech, respectively (see Section 3.2); and (c) we perform the two-stage training for EL2SP with SPD and the original dataset (see Section 3.3). Further details are fleshed out in the following subsections. 

\subsection{Pretraining of sequence-to-sequence model}
\label{3.1}
We adopt the transformer-based TTS pretraining for building a one-to-one VTN. The TTS model is similar to the VC model in decoding hidden representations into speech, while only a single speaker's dataset and corresponding transcriptions are needed. This naturally motivates us to transfer the ability from the TTS to the VC during pretraining stages. Fig. \ref{Fig. 1} (a) depicts the specific architecture containing two pretraining stages.

Given a large normal TTS corpus, $D_{TTS}$=\{$T_{TTS}$, $S_{TTS}$\}, where $T_{TTS}$ and $S_{TTS}$ denote the text and speech dataset respectively. During the decoder pretraining stage, we use $D_{TTS}$ for typical TTS training. Encoder can effectively produce the hidden representations of pure linguistic information of text, thanks to which the decoder has a robust capability of capturing the relation between various speech features and linguistic information. In the encoder pretraining stage, an autoencoder is involved in training with a reconstruction loss instead of the original TTS encoder part, while freezing the pretrained decoder. Here, the $S_{TTS}$ is used as both inputs and target dataset. In this manner, the encoder can be forced to accurately learn to extract analogous linguistic hidden representations from the speech instead of the text, provided that the decoder maintains the ability to recognize linguistic information from the encoded text. 

The above is the complete process of VTN pretraining. Note that we choose normal human corpus for pretraining because it can be easily obtained, and the goal for both EL2SP and normal VC is to convert the hidden linguistic representations to normal speech, thus enabling the EL2SP to inherit the useful capability. After that, fine-tuning stage can be implemented using a EL2SP corpus different from the $D_{TTS}$.

\subsection{Synthetic parallel data generation}
\label{3.2}
As illustrated in Fig. \ref{Fig. 1} (b), this process aims to produce EL and normal SPD from the TTS systems. Therefore, we train not only TTS for normal speech generation but also for EL speech generation. However, the initial source EL corpus, $D_{EL}$=\{$T_{EL}$, $S_{EL}$\}, and the target normal corpus, $D_N$=\{$T_{N}$, $S_{N}$\}, are too small to be trained from scratch. We perform a fine-tuning with the TTS model obtained by the decoder pretraining stage based on the two corpus, respectively, obtaining the corresponding models, $TTS_{EL}$ and $TTS_{N}$. Note that the performances of $TTS_{EL}$ and $TTS_{N}$ are still poor due to the original dataset being too small. Finally, we input an external set of text to the two systems simultaneously to generate low-quality EL and normal SPD, denoted by $S_{EL}^{SPD}$ and $S_{N}^{SPD}$, respectively.

\subsection{Two-stage EL2SP training}
\label{3.3}
As presented in Fig. \ref{Fig. 1} (c), the parameters of $VTN_{pt}$ from Section \ref{3.1}, can be used as valid \textit{a priori} information to initialize the VTN-based EL2SP model, thus achieving better performance and extremely efficient convergence compared to training from scratch. Hereafter, SPD and original natural speech pairs are concurrently pooled together as the training set to obtain the EL2SP model, $VTN_{1st}$, which is denoted as Stage I. Although SPD can model the intonation and speaker identity of normal speech and EL speech, it contains some misinformation due to the limited performance of the TTS models, which may inhibit EL2SP from building perfect hidden representations. Furthermore, we hypothesize that the limited natural dataset can provide a correction for training. So, stage II is set up to obtain the final EL2SP model, $VTN_{2nd}$, during which we initialize the parameters of $VTN_{1st}$ and re-input the original natural data pairs for training.

\section{EXPERIMENTAL evaluation}
\label{4}

\subsection{Datasets and implementation}
We built two original small-scale Japanese datasets to evaluate our proposed method. In one, a healthy male speaker recorded parallel utterances using an electrolarynx and his normal voice, denoted by SimuEL and NormSP, respectively. Each pair contained 413 sentences, totaling to 20 minutes. The other EL set came from a real male laryngectomee, denoted as RealEL. Since we could not record the normal corpus from this laryngectomee, we shared NormSP as the reference target speech. Note that RealEL contains only 200 sentences within 10 minutes, whose utterance contents are only part of NormSP, so this dataset belongs to a semi-parallel corpus.

Two experiments were designed based on this two datasets. For case 1, our goal was to convert SimuEL to NormSP (SimuEL-NormSP), which can simulate the normal corpus of patient was available. For case 2, we contrived to convert RealEL to NormSP (RealEL-NormSP), which was a more severe but practical case. Here, we used 20 utterances from the respective dataset as a development set, another 40 as an evaluation set, and the rest were used as a training set. To address the non-parallel part in RealEL-NormSP, we first supplemented the corresponding synthetic EL speech to utilize all feasible data for our method. Note that due to the datasets provided by two cases were different, for the sake of simplicity, we built different evaluation sets.

We used ESPnet \cite{18} to implement the proposed system, where all the speech signals were re-sampled at 24khz, and the 80-dimensional mel filterbanks with 2048 FFT points and a 300-point frame-shift were used to extract the acoustic features. We followed Section \ref{3.1} to build the pretrained VTN using the Japanese corpus named JSUT database \cite{19}, which contains 7696 utterances recorded by a female speaker, roughly 10 hours long. In addition, text data of JSUT database was also chosen for generating external SPD. VTN for constructing EL2SP and the TTS models for generating SPD have similar base configurations, using six-layer blocks with four attention heads in the encoder and decoder. The input feature sequence of the TTS model contains phoneme and pause information. For the waveform synthesis module, we used a Parallel WaveGAN (PWG) neural vocoder \cite{20} to reconstruct SPD and the output of EL2SP. The corresponding speaker-dependent PWG vocoders were trained separately using SimuEL, NormSP, and RealEL. Note that the PWG vocoder for NormSP was trained from scratch, while for SimuEL and RealEL were pretrained from an EL dataset with 1000 utterances. Moreover, we set up typical Transformer-based baseline models for case 1 and case 2. Their framework and pretraining process are identical to the proposed system but use only original parallel pairs without SPD during EL2SP training, i.e., 353 for case 1 and 140 for case 2.

\begin{table*}[ht]\small
\begin{minipage}[b]{0.5\textwidth}
\centering
\caption{Comparison results of the proposed method and\\ the baseline system for case 1: SimuEL-NormSP.}
\setlength{\tabcolsep}{0.2mm}{
\begin{tabular}{ccccccc}
\Xhline{1pt}
\textbf{Systems} & \multicolumn{2}{c}{\textbf{\begin{tabular}[c]{@{}c@{}}External SPD\\ Quality\end{tabular}}} & \multicolumn{4}{c}{\textbf{EL2SP Results}} \\ \Xhline{1pt}
\multirow{3}{*}{\begin{tabular}[c]{@{}c@{}}Proposed methods \\ using external SPD\\  with different datasizes\end{tabular}} & \multirow{2}{*}{EL} & \multirow{2}{*}{Normal} & \multicolumn{2}{c}{\multirow{2}{*}{Stage I}} & \multicolumn{2}{c}{\multirow{2}{*}{Stage II}} \\
 &  &  & \multicolumn{2}{c}{} & \multicolumn{2}{c}{} \\ \cline{2-7} 
 & CER & CER & MCD & CER & MCD & CER \\ \hline
SPD-1000 & 35.3 & 16.4 & 5.95 & 42.1 & 5.93 & 39.4 \\ \hline
SPD-2000 & 39.4 & 20.0 & 5.90 & 40.7 & 5.82 & 37.8 \\ \hline
SPD-4000 & 44.5 & 25.6 & 5.87 & 37.4 & 5.78 & 36.9 \\ \hline
SPD-7296 & 53.0 & 34.8 & 5.78 & 36.3 & 5.77 & 34.7 \\ \Xhline{1pt}
\multirow{2}{*}{Baseline} & \multirow{2}{*}{-} & \multirow{2}{*}{-} & \multicolumn{2}{c}{MCD} & \multicolumn{2}{c}{CER} \\ \cline{4-7} 
 &  &  & \multicolumn{2}{c}{6.37} & \multicolumn{2}{c}{51.4} \\ \Xhline{1pt}
\end{tabular}}
\label{TABLE:I}
\end{minipage}
\begin{minipage}[b]{0.5\textwidth}
\centering
\caption{Comparison results of the proposed method and\\ the baseline system for case 2: RealEL-NormSP.}
\setlength{\tabcolsep}{0.2mm}{
\begin{tabular}{ccccccc}
\Xhline{1pt}
\textbf{Systems} & \multicolumn{2}{c}{\textbf{\begin{tabular}[c]{@{}c@{}}External SPD\\ Quality\end{tabular}}} & \multicolumn{4}{c}{\textbf{EL2SP Results}} \\ \Xhline{1pt}
\multirow{3}{*}{\begin{tabular}[c]{@{}c@{}}Proposed methods \\ using external SPD\\ with different datasizes\end{tabular}} & \multirow{2}{*}{EL} & \multirow{2}{*}{Normal} & \multicolumn{2}{c}{\multirow{2}{*}{Stage I}} & \multicolumn{2}{c}{\multirow{2}{*}{Stage II}} \\
 &  &  & \multicolumn{2}{c}{} & \multicolumn{2}{c}{} \\ \cline{2-7} 
 & CER & CER & MCD & CER & MCD & CER \\ \hline
SPD-1000 & 49.5 & 17.2 & 6.66 & 26.9 & 6.59 & 24.5 \\ \hline
SPD-2000 & 54.8 & 21.0 & 6.48 & 27.7 & 6.41 & 25.3 \\ \hline
SPD-4000 & 61.5 & 25.7 & 6.28 & 26.3 & 6.18 & 21.9 \\ \hline
SPD-7296 & 71.7 & 33.7 & 6.26 & 29.1 & 6.24 & 23.3 \\ \Xhline{1pt}
\multirow{2}{*}{Baseline} & \multirow{2}{*}{-} & \multirow{2}{*}{-} & \multicolumn{2}{c}{MCD} & \multicolumn{2}{c}{CER} \\ \cline{4-7} 
 &  &  & \multicolumn{2}{c}{7.17} & \multicolumn{2}{c}{41.3} \\ \Xhline{1pt}
\end{tabular}}
\label{TABLE:II}
\end{minipage}
\end{table*}

\subsection{Evaluation metrics}
We performed objective evaluations using two several metrics, 1) mel cepstrum distortion (MCD), which was calculated between the ground truth samples and the generated samples to measure the spectral envelope distortion, and 2) the character error rate (CER) to evaluate the intelligibility. Meanwhile, as the quality of SPD needed to be examined in this study, we used CER as a rating criterion for SPD. Here, the ASR engine used to calculate CER was conformer-based architecture \cite{21} pretrained from Japanese laboroTV database \cite{22} and fine-tuned as in \cite{23}. It could recognize all original EL and normal speech with the best CER (in \%) of 14.7 and 6.9, respectively, which can be regarded as the upper bounds of ASR results to reflect the quality of SPD. 

We also performed subjective tests to evaluate the perceptual performance of VC on the generated speech in terms of naturalness. An opinion test to assess naturalness was conducted in terms of mean opinion score (MOS). Listeners were asked to rate the naturalness of each given speech (in the scale of one to five). For case 1 and case 2, twenty random utterances were chosen for the naturalness tests, respectively. More than thirty listeners who can speak Japanese were recruited. Audio samples are available online\footnote{\href{https://silenticymoon.github.io/EL2SP-demos/}{\scriptsize\url{https://forpapersub.github.io/For_SLT_2022/}}}.

\subsection{Experimental results}
\subsubsection{Objective evaluation}
\label{4.3.1}
\begin{figure}[!t]
\begin{minipage}[t]{1\linewidth}
  \centering
  \includegraphics[width=8.3cm]{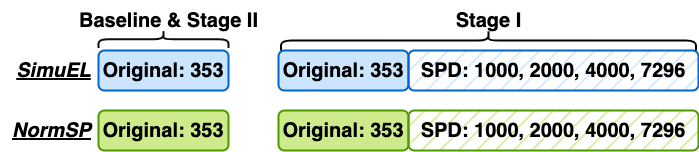}
  \centerline{\hfill\small(a) Case 1: SimuEL-NormSP\hfill}\medskip
\end{minipage}
\begin{minipage}[t]{1\linewidth}
  \centering
  \includegraphics[width=8.3cm]{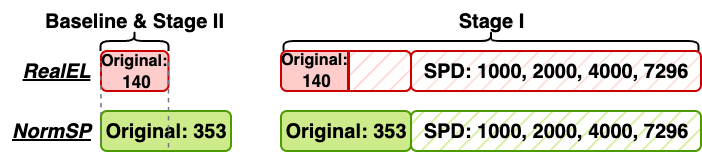}
  \centerline{\hfill\small(b) Case 2: RealEL-NormSP\hfill}\medskip
\end{minipage}
\caption{The training datasets for the baseline system, stage I and stage II of proposed method for case 1 and case 2, respectively. 1000, 2000, 4000 and 7296 represent the screened datasizes of SPD used for the experiments.}
\label{Fig. 2}
\end{figure}

To investigate the impact of SPD on EL2SP performance, we set up different amount of SPD for comparisons. Here, the term $datasize$ is used to represent the amount of the data. Note that the TTS models used to generate SPD were not universal due to the different training sets between case 1 and case 2. We built a filtering system on the ASR model to select SPD training pairs with better quality based on CER, so as to reduce the error for training. The filtering datasize was gradually increased from 1000 to use all available external SPD (7296), and performed the two-stage training procedure stated in Section \ref{3.3}. Fig. \ref{Fig. 2} indicates the specific training datasets of the baseline and the proposed method for both cases. 

The objective evaluation results of two-stage method for case 1 and case 2 with different SPD datasizes are listed in Table \ref{TABLE:I} and Table \ref{TABLE:II}, and the CER values of SPD are attached as a demonstration of reliability to indicate the quality of SPD.

For case 1, the CER values characterizing the quality of EL and normal SPD show a steady increase from 35.3 and 16.4 to 53.0 and 34.8, respectively. Nevertheless, the outcomes are consistently improved in both MCD and CER when the SPD is increased from 1000 to 7296 in stage I. In addition, all the results using SPD are significantly better in terms of quality and intelligibility than the baseline system. During stage II, we conducted the same training procedure for the models obtained from stage I with different SPD datasizes, which yielded better results. It can be determined that in case 1, the proposed two-stage training using SPD and original dataset is able to assist the system to build the rich-linguistic hidden representations from stage I.

In case 2, although normal SPD keeps the same quality level as in case 1, the quality of EL SPD in Table \ref{TABLE:II} is much inferior to that in Table \ref{TABLE:I}. Since the datasize of the original training corpus for RealEL (140) is over half less than that for SimuEL (353), leading to worse performance of the EL TTS model. Also, the CER of EL SPD in Table \ref{TABLE:II} has a sharp growth from 49.5 to 71.7. Surprisingly, the systems in stage I still outperform the baseline system with a notable margin in terms MCD and CER even after using SPD. Here, the MCD continuously decreases as SPD datasize increases from 1000 to 7296. Furthermore, the optimization of CER is not obvious and remains at a steady level. When all SPD (7296) is pooled into training, the MCD value turns out to be the best, while the CER increases compared with the scenario of 4000. We expect that such large-scale and low intelligible SPD would improve the speech quality while affecting the accuracy of utterance contents. Next, the results of stage II are all better than stage I for each datasize, especially when more SPD with lower quality is involved in stage I, e.g., CER decreases by 5.1 on average as SPD is 4000 and 7296.
\begin{figure}[!t]
\setlength{\abovecaptionskip}{0cm}
\begin{minipage}[b]{0.48\linewidth}
  \centering
  \centerline{\includegraphics[width=44mm]{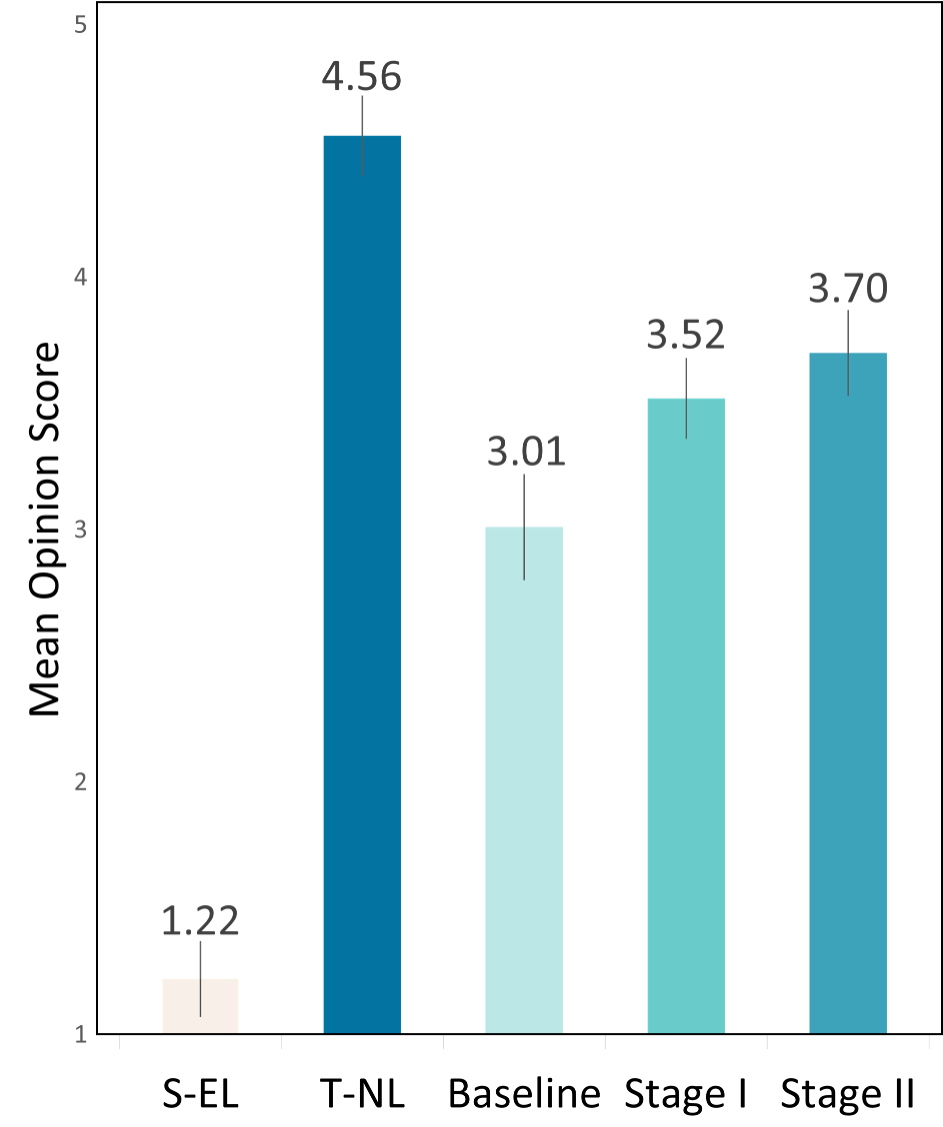}}
  \centerline{\hfill\small(a) Case 1: SimuEL-NormSP \hfill}\medskip
\end{minipage}
\hfill
\begin{minipage}[b]{0.48\linewidth}
  \centering
  \centerline{\includegraphics[width=44mm]{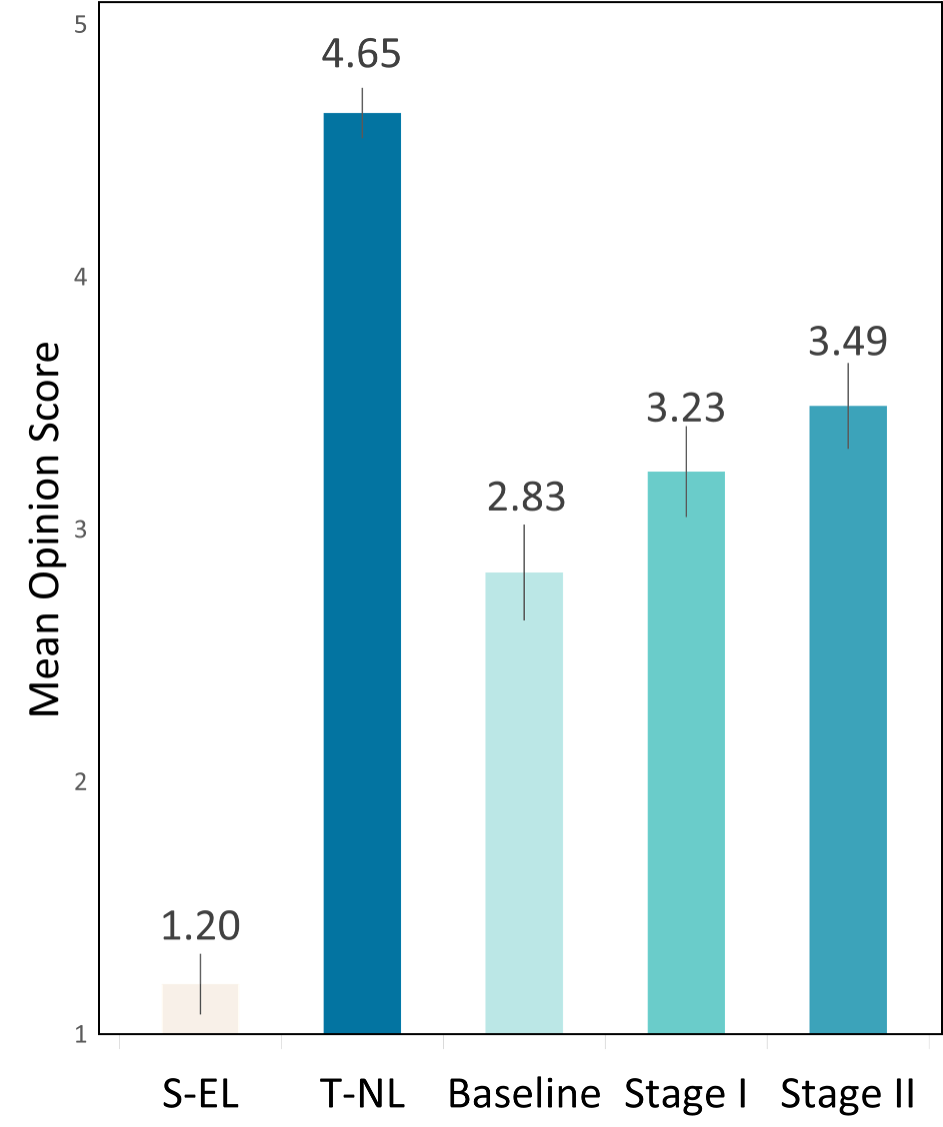}}
  \centerline{\hfill\small(b) Case 2: RealEL-NormSP\hfill}\medskip
\end{minipage}
\caption{MOS results with 95\% confidence intervals for naturalness. S-EL and T-NL indicate source EL and target normal, respectively.}
\label{Fig. 3}
\end{figure}

The overall results demonstrate robustness of the proposed method. i.e., the low-quality SPD allows the system to obtain a better performance than the baseline in stage I. And stage II can maximize the advantages of large-scale SPD, consequently having negligible negative impact on CER.

\subsubsection{Subjective evaluation}
The outcomes of SPD datasize of 4000 were randomly selected for subjective evaluation tests for both cases. Fig. \ref{Fig. 3} illustrates the results in which five types of speech were mixed into the test set. It could be observed that the seq2seq-based baseline systems return reasonable results  ($\approx$3) in terms of naturalness, even though the intelligibility is unideal. It shows that the listeners paid more attention to the factors such as fluency and voice quality rather than the linguistic contents. Moreover, the proposed method can yield significantly better performance from stage I compared to the baseline system. Meanwhile, although case 2 has a more difficult mapping process between different speakers, its results are only slightly worse than those of case 1, showing the effectiveness of the proposed method. In addition, the results of both cases from stage I to stage II reveal a consistent and stable improvement, which is similar to the findings in Section \ref{4.3.1}.

\section{Conclusion}
\label{sec:majhead}

We demonstrate the proposed two-stage method incorporated with large-size low-quality SPD is a simple and effective optimized EL2SP compared to the baseline system. In particular, the effectiveness and robustness of low-quality SPD have been verified, and the system performance is positively correlated with the increase of SPD. Moreover, the second stage training can further diminish the negative impact of SPD and optimize the results. Interestingly, the texts that generate external SPD are from the VTN pretraining database. This confirms the high cost-effectiveness of our method. Current progress opens the path for future studies including: 1) randomly selecting SPD to design experiments to clarify the influence of the datasize and the quality of SPD on the results; and 2) re-establishing a semi-parallel dataset with less normal speech to investigate its effects on our proposed method.

\section{Acknowledgement}
\label{sec:print}
This work was partly supported by JST CREST JPMJCR19A3, and AMED JP21dk0310114, Japan. 



\end{document}